\documentclass[11pt]{article} 

\usepackage[utf8]{inputenc} 

\usepackage{todonotes}
\usepackage{amsmath}
\usepackage{xr-hyper} 
\usepackage{hyperref} 
\usepackage{hyperref}
\usepackage{amssymb}
\usepackage{natbib}
\usepackage{float}
\usepackage{booktabs}
\usepackage{multirow}
\usepackage{url}
\usepackage{xurl}

\usepackage{geometry} 
\usepackage{graphicx} 

\usepackage[parfill]{parskip} 

\usepackage{booktabs} 
\usepackage{array} 
\usepackage{paralist} 
\usepackage{verbatim} 
\usepackage{subfig} 

\usepackage{fancyhdr} 
\usepackage{sectsty}
\allsectionsfont{\sffamily\mdseries\upshape} 

\usepackage[nottoc,notlof,notlot]{tocbibind}
\usepackage[titles,subfigure]{tocloft}

\usepackage{authblk}

\title{Reconstructing temporal multi-relational firm networks at scale using large language models. The case of the semiconductor industry}

\author[1,2,3]{Şeyda Köse}
\author[2,4,5]{Christian Diem}
\author[1,2,3]{Elma Dervic}
\author[1,7]{Klaus Friesenbichler}
\author[1,2]{Georg Heiler}
\author[2]{Jan Hurt}
\author[1]{Hernan Picatto}
\author[1,2,3]{Peter Klimek}

\affil[1]{Supply Chain Intelligence Institute Austria (ASCII), Vienna, Austria}
\affil[2]{Complexity Science Hub Vienna, Vienna, Austria}
\affil[3]{Medical University of Vienna, Section for Science of Complex Systems, CeMSIIS, Vienna, Austria}
\affil[4]{Smith School of Enterprise and Environment, University of Oxford, Oxford, United Kingdom}
\affil[5]{Institute of New Economic Thinking, University of Oxford, Oxford, United Kingdom}
\affil[6]{Austrian Institute of Economic Research, Vienna, Austria}

\date{}

\begin{document}
\maketitle

\begin{abstract}
The semiconductor industry is foundational to modern technology, yet its complex global multi-relational firm network remains poorly understood, posing challenges to scientists, firms and policymakers.
Traditional analysis relies on proprietary databases that are often expensive, incomplete, and slowly updated, limiting their ability to capture rapidly evolving dependencies. Here, we demonstrate that a novel, generalizable methodology combining Large Language Models (LLMs) with open web data can reconstruct this network and its structural dynamics at scale.
We identify and classify supply-chain, partnership, and ownership links from 170 million semiconductor firm webpages, yielding a temporal network of over 1,300 linked firms. We validate link-extraction quality (Precision: 0.884; F1-score: 0.784), network overlap and complementarity with a proprietary database, and consistency with aggregate economic data. 
Our network reveals a temporary 9\% decline in edges during the 2022 chip shortage, rapid increases in the centrality of AI supply-chain bottleneck firms such as NVIDIA, and geographic realignment of interfirm relations amid geopolitical turbulence.
This generalizable framework overcomes barriers to transparency and provides essential, up-to-date maps for assessing resilience and informing policy across strategically relevant sectors.

\end{abstract}

\section{Introduction}

Global economic activities are organized through complex webs of relationships inter-linking almost all firms. While often modeled strictly as production networks \cite{pichler2023sci}, where edges represent flows of goods or services, these systems are in reality \textit{multi-relational firm-to-firm networks} that encompass not only supply-chain links but also strategic partnerships, technology collaborations, credit relations and complex ownership structures. The structure of such uni-relational networks influences macroeconomic resilience \cite{acemoglu2012network, cimini2015systemic, klimek2019resil}, the spreading of economic shocks \cite{inoue2019firm, acemoglu2011, carvalho21, diem2024, Chakraborty2024, papadopoulos2025climate}, economic systemic risk \cite{diem2022quantifying, mancini2025evolution} food supply security \cite{Diem03052025}, economic growth \cite{McNerney2022}, corporate control dependencies \cite{vitali2011network, mizuno2020power} or the pace and direction of innovation \cite{dahlke2024epidemic,zhu2006innovation}. When interlinkages between firms span multiple relationship types simultaneously -- e.g., the planned 100 billion dollar investment of NVIDIA in its major customer OpenAI to finance the purchase of GPUs -- concerns about economic systemic risks increase \citep{economist_nvidia_openai_2025}. Understanding these risks requires gathering and analyzing firm-to-firm networks in a multi-relational way. In sectors characterized by high levels of specialization and global fragmentation, such as semiconductors, mapping this multi-relational topology of interdependence is especially critical. Semiconductor manufacturing involves distinct and geographically dispersed stages, including chip design, fabrication, assembly, tool \& equipment production, and testing, frequently distributed across countries like the United States, Taiwan, China, South Korea, Japan and the Netherlands \cite{ASCII2024, oecd2025mappingsemiconductor}. Albeit these networks are important, publicly available firm-level data on their structure remains limited and fragmented \cite{pichler2023sci, bacilieri2023firm}.

Existing efforts to map these economic architectures often rely on sector-aggregated datasets such as the OECD’s Inter-Country Input-Output (ICIO) tables, World Input-Output Database (WIOD) \cite{Dietzenbacher01032013}, or CEPII’s BACI trade database. While valuable for macroeconomic modeling, these sources only offer coarse-grained, industry-level information and can introduce aggregation bias that distorts the modeling of important economic dynamics like shock propagation \cite{diem2024}. In contrast, firm-level data provides much finer granularity and allows for developing a new type of economic models \cite{pichler2023sci} but is subject to serious limitations. Administrative datasets such as VAT transaction records from Ecuador, Belgium and Hungary offer high-quality firm-level information \cite{dhyne2015belgian, diem2022quantifying, mungo2023reconstructing, bacilieri2023firm}, but they are usually hard to access and limited to national scopes, which cannot capture the structure of global value chains. Commercial datasets like FactSet or S\&P Capital IQ offer a wider geographic scope but are prohibitively expensive and often limited to large, publicly listed firms, covering only a narrow slice of the global economy.

Despite growing interest in the structure of firm-level networks, no large-scale, publicly available dataset currently exists to capture these relationships at a global level \cite{maccarthy2022mapping}. This data gap is particularly concerning in the case of semiconductors, which are the basis for a vast array of modern technologies, from smartphones and electric vehicles to military systems and artificial intelligence. The COVID-19 pandemic and the following chip shortages \cite{wu2021analysis,mohammad2022global} illustrated how disruptions in semiconductor supply chains can trigger cascading effects across entire economies, from delayed auto production in the U.S. to inflationary pressures on consumer electronics globally. In recent years, governments have therefore repeatedly used central roles of their domestic semiconductor companies to further their political aims \cite{bednarski2025geopolitical}. Given these systemic risks and geopolitical sensitivities, accurately mapping the multi-relational semiconductor network at the firm-level is essential for anticipating choke points, assessing economic inter-dependencies, and informing industrial or trade policy.

To bridge this gap, recent research has increasingly turned to mathematical and computational methods for reconstructing production and supply networks. These methodologies generally fall into two categories \cite{mungo2024reconstructing}. (i) Inferring network topology from partial economic data, using probabilistic models, maximum entropy principles, or machine learning to predict missing links \cite{mungo2023reconstructing, ialongo2022reconstructing}. Yet, as these methods only estimate the likelihood that a link exists, they may recover overall network topology but cannot reliably quantify dependencies between specific firms, limiting their use for policy decisions. (ii) Extracting explicit relationships from unstructured text. Wichmann et al.\cite{wichmann2020extracting}, for instance, utilized deep neural networks to mine firm-to-firm connections from news articles, while recent contributions have developed frameworks leveraging Knowledge Graphs and Large Language Models (LLMs) to enhance supply chain visibility using public unstructured text, such as Wikipedia \cite{almahri2024enhancing} or SEC 10-K filings \cite{jackson2025supply}. However, these approaches predominantly rely on third-party narratives, such as news coverage, standardized reports or Wikipedia entries, which inherently limit visibility to prominent, publicized entities. This reliance overlooks the vast majority of less publicized, yet economically important, business-to-business relationships that are only observable through direct firm disclosures. Hence, there is a significant lack of approaches that leverage first-party data for scalable, accurate information extraction without task-specific model training.

In this paper, we develop a novel, scalable framework for reconstructing the global multi-relational firm-to-firm network by combining firm-disclosed information with Large Language Models (LLMs) and apply it to the semiconductor industry. Our approach begins by screening a broad candidate pool of over 21,000 semiconductor-related firms identified via business intelligence data (ORBIS). We then retrieve the historical web presence of these firms from the Common Crawl archive (2015–2025) to isolate the subset of companies with active, discoverable business relationships. By processing the textual content of these archived websites with a pretrained LLM (GPT-4o-mini), we automatically extract and classify directed inter-firm linkages, distinguishing between \emph{supply} relationships, strategic \emph{partnerships}, and firm \emph{ownership} links. This process yields a reconstructed temporal network of over 1,300 linked firms, offering an open, low-cost complement to traditional commercial business intelligence data sources. We validate this network through human annotation, comparison with macroeconomic trade and production data (OECD ICIO and BACI), and a statistical benchmark against S\&P Capital IQ, where we employ null models to quantify the significance of the observed network overlap. In the following section, we present the structural evolution of this network, demonstrating its capacity to capture critical industry dynamics, from the disruptions caused by the COVID-19 pandemic to the rise of AI-related companies as central hubs.

\section{Results}

\subsection{Network descriptives}

\begin{figure}[tbp]
    \centering
    \includegraphics[width=\textwidth]{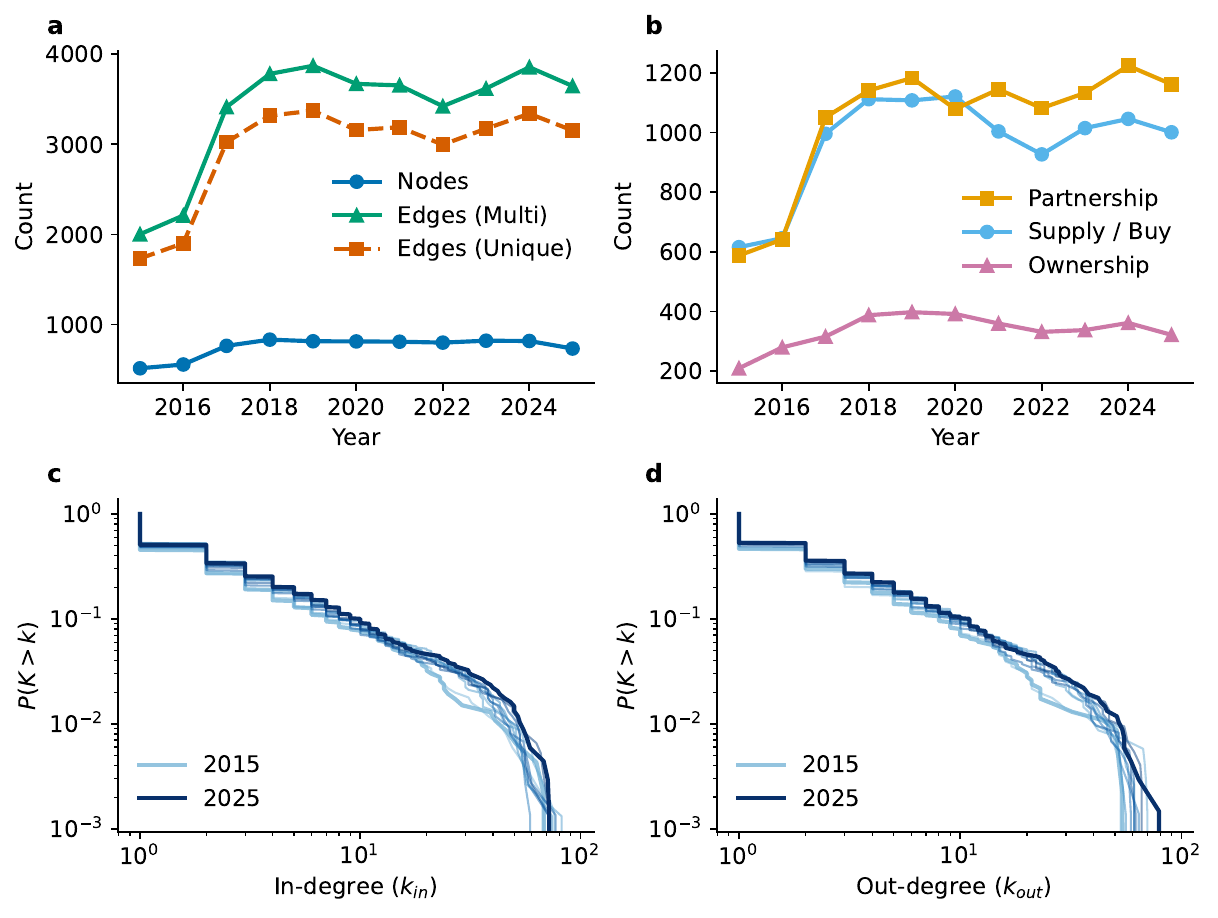}\caption{\textbf{Evolution and structure of the reconstructed firm-level production network (2015–2025).}
    \textbf{a} Evolution of network size, showing the number of nodes (blue dots), the count of multi-relational edges, where different relationship types between the same firm pair are counted separately (green triangles) and the count of unique directed relationships, where multiple relationship types between the same firm pair are collapsed into one link (orange squares).
    \textbf{b} Composition of the network by edge type, distinguishing between supply/buy links (blue dots), partnerships (yellow squares), and ownership (green triangles) relations.
    \textbf{c,d} Complementary cumulative distribution functions (CCDFs) of the collapsed network of in-degree ($k_{in}$) and out-degree ($k_{out}$), respectively, shown on log--log scales.
    In panels \textbf{c} and \textbf{d}, lighter lines correspond to earlier years, while darker lines highlight the most recent observation periods.}
    \label{fig:network_descriptives}
\end{figure}

Figure~\ref{fig:network_descriptives} summarizes the evolution and structural properties of the reconstructed multi-relational firm-level network between 2015 and 2025. The number of active nodes (companies) in each yearly snapshot of the weakly connected component of the network grows from approximately 500 in 2015 to nearly 820 in 2024. Over the same period, projecting all relationship types on a single-layer network -- where two nodes are connected if they have a relationship of any type --, the total number of links increases from 1,733 to 3,340. When counting all recorded relationships separately, the count rises from 2,002 to 3,856 links. We attribute the relatively small network size in 2015–2016 to incomplete coverage in early Common Crawl snapshots  (not a genuine absence of inter-firm links), and, hence, use only the snapshots from 2017 onward for our analysis.

The number of unique edges peaks in 2019 (3,373 unique edges), followed immediately by a contraction in 2020 (3,158 edges) that further decreases to a low point in 2022 (2,996 edges), before increasing again until 2024. This timing corresponds with the onset of the COVID-19 pandemic (coming into effect in 2020) and the complex supply chain shocks that followed (2021-2022). Although demand for semiconductors dropped significantly during the early stages of the pandemic, it recovered rapidly following a 2020 boom in consumer electronics spending; however, the supply chain failed to expand capacity quickly enough to keep pace throughout 2021 and 2022.

Figure~\ref{fig:network_descriptives}b distinguishes between three types of relationships (\emph{supply-chain}, \emph{partnership}, and \emph{ownership}) and shows their evolution over time. In in Fig. 1a, the drop of firm relations between 2021 and 2022 is mostly driven by decreases in supply chain links. Note that while in our directed network model (Fig.~\ref{fig:network_descriptives}a), partnerships are treated as bidirectional links, Figure~\ref{fig:network_descriptives}b reports unique undirected pairs to avoid double-counting these reciprocal interactions. 

The bottom panels display the complementary cumulative distribution functions (CCDFs) of firms' in-degree (number of buy-links) and out-degree (number of supply-links), respectively, for the single-layer network that collapses all relationship types. Both distributions are heavily right-skewed. This indicates that while the majority of firms maintain only few links (partnerships, supply-chain and ownership links) a small subset of firms acts as highly connected hubs with a disproportionately large number of connections, a structure consistent with findings from empirical studies of supply networks \cite{fujiwara2010large, bacilieri2023firm} and ownership networks \cite{vitali2011network}.

\subsection{Validation}

\subsubsection{LLM based link-extraction validation}

To ensure the reliability of our automated extraction pipeline, we evaluate the performance of the GPT-4o-mini-based model against a ground-truth dataset of human-annotated webpages. We assessed the model's accuracy on two distinct levels: \textbf{binary classification}, which measures the ability to correctly distinguish between pages containing relevant business relationships and those that do not; and \textbf{multi-class classification}, which tests the model's ability to correctly categorize the different types of relationships (i.e., supplier, partner, or owner) and its direction.

The performance metrics for these tasks are summarized in Table~\ref{tab:gpt_validation}. Very high precision scores across both tasks indicate that the model generates few false positives, ensuring that the reconstructed network edges represent high-confidence relationships. Notably, the multi-class classification precision (0.918) exceeds that of binary classification (0.884). This suggests that once the presence of a link is established, identifying the specific relationship type from the context is relatively robust, whereas detecting the initial existence of a link remains the more difficult task due to the unstructured nature of web content.  The high precision values stem from a conservative calibration of the link retrieval pipeline, that in turn causes lower levels of Recall of 0.704 and 0.648, respectively.

\begin{table}[ht]
    \centering
    \caption{Performance of GPT-based relationship classification.}
    \label{tab:gpt_validation}
    \begin{tabular}{lcc}
        \toprule
        \textbf{Metric} & \textbf{Binary classification} & \textbf{Multi-class classification} \\
        \midrule
        Accuracy  & 0.790 & 0.648 \\
        Precision & 0.884 & 0.918 \\
        Recall    & 0.704 & 0.648 \\
        F1-score  & 0.784 & 0.758 \\
        \bottomrule
    \end{tabular}
\end{table}

\subsubsection{Validation against aggregate economic data}   

To assess whether the reconstructed firm-level network recovers the established macroeconomic trade patterns, we validate it against the OECD ICIO and BACI international trade database. When comparing log-transformed trade volumes (excluding domestic flows to focus strictly on international trade), our reconstructed network exhibits a robust alignment with official benchmarks. Firm-level edges are aggregated to the country level using a gravity-weighted scheme based on revenue shares, and domestic flows are excluded throughout. Comparing log-transformed bilateral trade volumes of each datasets, we find a Pearson correlation of 0.64 and a Spearman rank correlation of 0.68, averaged across 2015–2022. These patterns are further corroborated by a parallel comparison against the BACI international trade database, which yields similarly strong agreement with an average log-linear Pearson correlation of 0.61 and a Spearman correlation of 0.70 over the same timeframe. Notably, both metrics in both comparisons remain stable across the full period, suggesting the reconstruction is consistent rather than period-specific.

\subsection{Comparison with a commercial supply chain dataset}

To assess how our newly constructed network compares with widely used commercial data, we benchmark it against the S\&P Capital IQ dataset, used in prior supply-chain network research, e.g., \cite{Chakraborty2024}. While both datasets are partial and subject to different coverage biases, S\&P serves as a curated commercial benchmark. We extract S\&P supplier–customer relationship data for semiconductor companies (see Methods section), and compare it against our supply-chain network, excluding partnerships and ownership ties.\footnote{S\&P Capital IQ covers 1996--2026 while our web-scraped data spans 2015--2025. Snapshots from a given crawl year may reflect relationships established well before that date. We therefore compare aggregated networks across the full available period of each dataset.}

After filtering both networks to directed supplier--customer links, the S\&P Capital IQ network contains 17{,}997 firms connected by 48{,}671 directed edges, while the reconstructed web-based network contains 999 firms and 3{,}407 directed edges. After matching the names across datasets, the two datasets share 1{,}469 S\&P entity records which map to 493 unique domains. The connectivity between these common firms is high in both datasets: S\&P reports 3{,}031 edges and our reconstructed network reports 1{,}692 edges. The overlap of exact edges is 380 directed supplier--customer links, i.e., about 22\% of edges we identify are also contained in the S\&P data. The high overlap is substantial given the vastly different ways of constructing the networks and shows that firms report supply-chain relationships on their homepages that are also picked up commercial business intelligence providers. Importantly the overlap also suggests that about 78\% of supply-chain links we infer from company homepages are not contained in commonly used proprietary data sets, showing that our reconstruction method makes an important contribution to mapping so far unchartered territory of the semiconductor industry supply-chain network. As shown by Fig. 1b, supply chain relations account for less than half of the overall relationships we identify, hence, our multi-relational network contains even more complementary information relevant to understanding the dynamics of the semiconductor industry.
Interestingly, the coverage of the two datasets differs across firms, e.g., for ARM there are more customers recorded in our data than in the S\&P data, while for Samsung we see the opposite, similarly Intel has more suppliers in our data and S\&P reports more suppliers for Samsung. For more details see supplementary Fig. \ref{fig:degree_comparison}. 

\begin{figure}[t]
    \centering
    \includegraphics[width=0.9\textwidth]{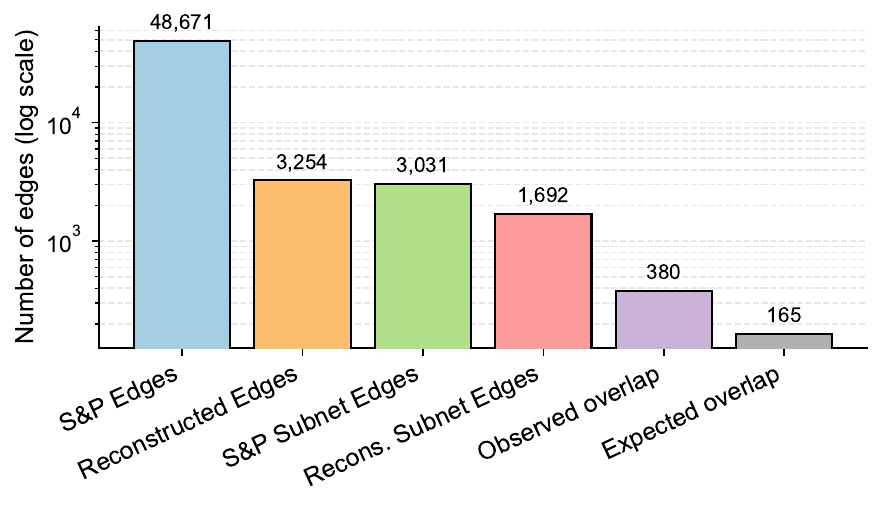} 
    \caption{\textbf{Comparison of our newly constructed supply-chain network with S\&P Capital IQ data.} The chart compares the total edge counts in the S\&P ground truth and the reconstructed network, the number of edges in the subnetworks induced by the overlapping nodes, alongside the observed and expected overlaps using the configuration model as suitable statistical null model.}
    \label{fig:sp_validation_combined}
\end{figure}

\paragraph{Statistical significance of the overlap.}
Because the two datasets originate from different incomplete sources, a simple count of overlapping edges between the two networks alone is not informative to assess their complementarity, as edges will theoretically overlap also by pure chance. Hence, the observed overlap must be evaluated relative to statistical expectations under suitable network theoretic null models. Here, we use the configuration model as a suitable baseline that preserves the degree sequence of both networks and thus controls for the trivial tendency of high-degree firms to appear in many relationships in both datasets (see Methods~\ref{conf_model} for formal derivations). We assess significance via Monte Carlo rewiring and further decompose the overlap by core-periphery network tiers using $k$-core decomposition, which allows us to test whether the agreement is concentrated among dominant hub firms or extends across the full network structure (see Supplementary Information for complete results).

The observed overlap of 380 directed edges is strongly significant relative 
to the configuration model null ($z = 23.14$, $p < 10^{-3}$), indicating 
that the two networks agree far more than can be explained by their degree 
sequences alone. Decomposing statistical significance by network tier reveals that this agreement is not confined to the most connected hub firms: all three tiers --core--core, core--periphery, and periphery--periphery -- are statistically significant ($z > 10$ in all groups), with high $z$-scores also among peripheral firms ($z = 15.01$).

\subsection{Temporal Network Analysis}
One of the key features of our reconstruction procedure is that we can build a coherent temporal network representation of the semiconductor firm-to-firm network at high observation frequency.
To understand the temporal dynamics of the global semiconductor firm-to-firm network, we analyze how the network evolves over time, particularly around major global disruptions. One such disruption is the global chip shortage (2020–2023). Triggered by pandemic-related lockdowns and a rapid shift in consumer demand, this period saw critical supply chain bottlenecks that fundamentally altered firm-to-firm dependencies.

The newly observed connections between firms show that the relative composition of international links fluctuates over the years, reflecting changes in the allocation of global supplier–customer relationships across blocs. In 2020, US-China connections dropped sharply just as China’s ties with the EU were expanding. This trend has intensified recently, with new links between U.S. and Chinese firms decreasing considerably. Detailed visualizations of new links and their country composition are provided in the Supplementary Information Figure~\ref{fig:si-new-links}.

\subsubsection{Regional Reorientation of Firms}

\begin{figure*}[t]
    \centering
    \includegraphics[width=\textwidth]{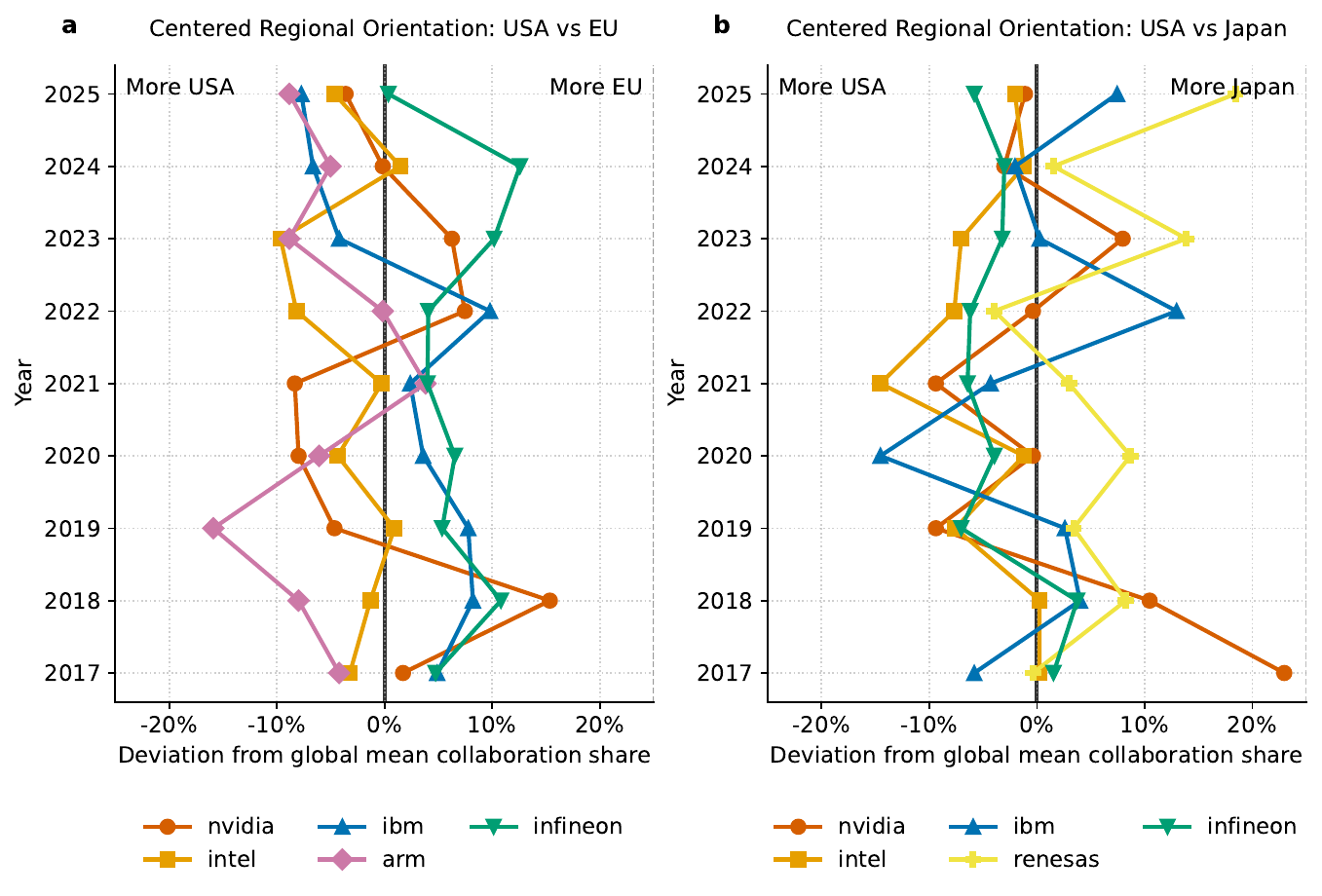}
    \caption{\textbf{Centered regional collaboration trajectories of selected firms (2017--2025).} Each line shows the temporal evolution of a firm’s relative regional orientation, measured as the share of edges with the focal region among its immediate (1-hop) neighbors in the reconstructed network and centered by the global mean over the full observation period. 
\textbf{a} United States versus European Union. 
\textbf{b} United States versus Japan.}

    \label{fig:regional_trajectories}
\end{figure*}

Beyond aggregate link trends, our framework allows us to track how individual companies adjust their regional alignment over time. Figure~\ref{fig:regional_trajectories} illustrates this by mapping the evolving position of major semiconductor firms between the United States, European Union, and Japan from 2017 to 2025. For each firm, we compute the annual share of its direct neighbors belonging to a specific region pair and center these values by subtracting the firm's own global mean share across the entire period. Overall, there is a trend towards stronger U.S. versus EU or Japanese integration of these companies over time, with some exceptions to this overall trend, such as Renesas, Intel, and IBM shifting towards Japan. 
Infineon maintains a consistent and strong orientation toward the European Union until 2024, whereas Intel leans slightly to the U.S. and IBM display more balanced profiles, fluctuating near the global average. ARM (UK) appears consistently more closely integrated with the U.S. network than with the EU throughout the observation period, reflecting its central role in design activities that are heavily concentrated within the U.S. ecosystem. NVIDIA, in contrast, exhibits a more volatile orientation. After a temporary realignment with European and Japanese partners peaking around 2023, the firm has shifted sharply back toward the United States in the most recent years (2024–2025), restoring a strong US-centric alignment. Interestingly, firms in Fig. 3a unanimously move their alignment from the EU to the U.S., which is a potential effect of the recent AI boom.

\begin{figure}[!ht]
    \centering
    \includegraphics[width=\textwidth]{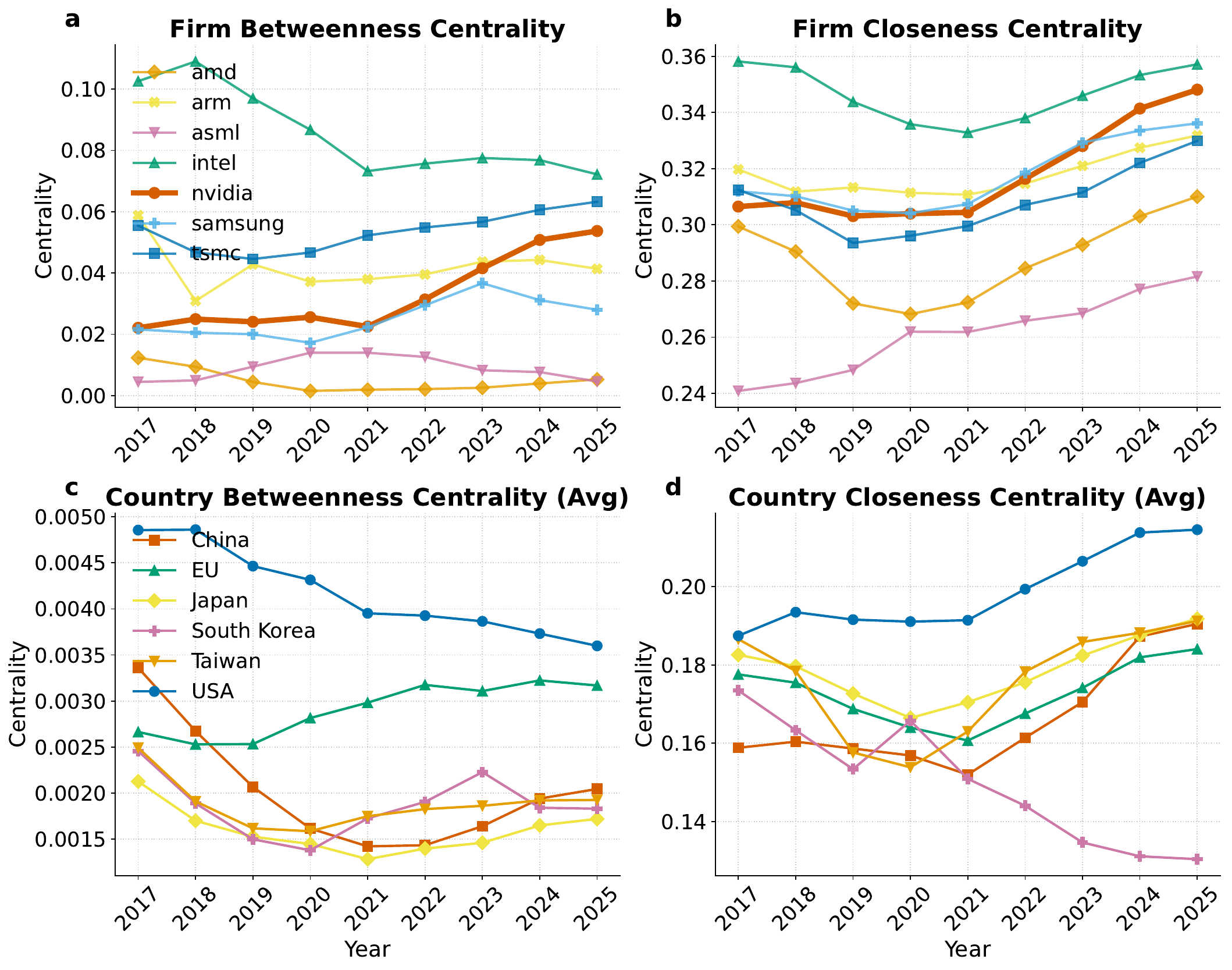}
    \caption{\textbf{Multi-scale evolution of network centrality (2015–2025).} \textbf{a-b} Firm-level dynamics: Evolution of \textbf{a} betweenness and \textbf{b} closeness centrality for selected leading firms (3-year rolling average). NVIDIA exhibits a sustained increase in both measures, contrasting with the relatively stable or declining trends of other major firms, consistent with its rising importance in AI-related supply chains. \textbf{c-d} Regional dynamics: Average \textbf{c} betweenness and \textbf{d} closeness centrality aggregated by region (arithmetic mean). In \textbf{c}, the U.S. maintains the highest brokerage role despite a gradual decline, while the EU exhibits a steady increase. In \textbf{d}, the U.S. shows a notable increase in closeness centrality from 2021 onwards, indicating a shift toward greater network centrality of US-based firms.}
    \label{fig:NVIDIA_centralities}
\end{figure}

\subsubsection{Emergence of AI}

Recent advancements in artificial intelligence (AI) have had a profound impact on the global semiconductor landscape. To investigate this shift in more detail, we examine how multi-relational network centrality for major firms evolves over time. We report betweenness and closeness centrality  -- particularly betweeness is a good indicator for how much of a bottle neck a company is within a supply chain \cite{brintrup2016topological}, while closeness captures how quickly a firm can reach, or be reached by, other firms through the network.\footnote{We computed multiple centrality indices and find that they exhibit qualitatively similar trends.} We find diverging trends in betweenness centrality across firms. NVIDIA shows a striking upward trend, particularly from 2022 on, the release data of ChatGPT and the start of the AI boom (see Figure~\ref{fig:NVIDIA_centralities}(a,b)). Similarly, the betweeness centrality of TSMC -- the key manufacturer of GPUs and the sole manufacturer for NVIDIA -- exhibits a continuous upward trend in betweeness since 2020. This suggests that our network tracks how NVIDIA and TSMC have both become an increasingly important intermediary in the firm network, likely due to the large increases of demand for GPUs used in deep learning, generative AI training and inference. In contrast more traditional and CPU focused firms such as Intel and ARM, show a clear downward trend.
Closeness centrality increases across most firms, particularly for NVIDIA, suggesting that these central firms are becoming more tightly interconnected in the network, possibly reflecting intensified coordination among major players. 

Heterogeneous firm dynamics are reflected in different trajectories across countries. Therefore, we calculate the average centrality score for all firms in a country. Overall, betweenness centrality decreased in most countries between 2015 and 2020 (Figure~\ref{fig:NVIDIA_centralities}c). After 2020, most countries exhibited relatively stable betweenness centralities, with the notable exceptions of China and the EU (increasing), while the U.S. betweeness centrality noticeably declined. Closeness centrality increased consistently for the US, partly driven by NVIDIA's rise (see Figure~\ref{fig:NVIDIA_centralities}d). 
Most other countries exhibit a `U-shaped curve', with a minimum around 2020.

\section{Discussion}

In this study, we present the first comprehensive approach to reconstructing multi-relational firm networks (supply-chain, ownership and partnership links) at scale based on self-declared company information retrieved from openly accessible web archives. Using the semiconductor industry as a case study, we demonstrate how large language models (LLMs) can be leveraged to transform unstructured online text into structured network representations, enabling the analysis of industry-wide dynamics with high temporal resolution.

Despite partial observability of the network, the extracted relationships are highly reliable. Prior work using neural network models for supply-chain links reports only moderate performance, particularly for directionality \cite{wichmann2020extracting}. More recent LLM and retrieval-based approaches for supply-chains improve precision but often suffer from low recall; for example, \cite{jackson2025supply} report precision around 0.83 with recall below 0.20. In contrast, our GPT-based approach for multi-relational links achieves both high precision and substantially higher recall (precision $>0.88$, recall $>0.70$; Table~\ref{tab:gpt_validation}), resulting in significantly stronger overall performance. Our method achieves even higher precision ($0.92$) for classifying edges into supply-chain, ownership and partnerships, while having a slightly lower recall ($0.65$). Remaining errors are largely due to genuinely ambiguous text, where even human annotators disagree. To explicitly address this ambiguity, our prompt was designed to be conservative: whenever a relationship could not be inferred unambiguously, the model was instructed to return no relationship. This prompting mitigates false positives while maintaining substantially higher coverage than prior methods. 

To assess the informational content of the reconstructed network, we compare it to an industry-standard commercial business intelligence dataset, which is itself incomplete and subject to distinct sampling biases. Rather than treating either dataset as ground truth, we evaluate whether they provide consistent views of the same underlying supply network and whether our method can extend existing data sources. We find that a substantial fraction of firms in the reconstructed network are also present in S\&P Capital IQ and that 22\% of the links we have identified are also reported in the S\&P data. This number of shared supplier--customer links is 2.3 times higher than expected under a suitable statistical null model (degree-preserving configuration model). Remarkably, the results point to a non-random alignment in the structure of inter-firm relationships captured independently by the two sources, which are based on very different data collection processes. Importantly, this agreement is not limited to the most prominent firms. When decomposing the network into core--core, core--periphery, and periphery--periphery interactions, we find statistically significant excess overlap in all categories. This suggests that the consistency between datasets extends beyond the network core and reflects broader structural features of the supply system. Taken together, these findings indicate that the web-based reconstruction recovers a substantial and structurally meaningful subset of real-world supply-chain relationships. Importantly, about 78\% of the links we identify are not present in the commercial database. Additionally, we detect also semiconductor firms that are not contained in the S\&P data. Due to our validation we can be confident that the large number of additional links provides a valuable contribution for creating better maps of the semiconductor industry ecosystem. 

The temporal resolution of our multi-relational firm-to-firm network reconstruction enables the study of salient industry-wide dynamics. We discuss how the new data set relates to key trends of the semiconductor industry over the last decade.

First, we illustrate this by examining temporal changes in network structure associated with supply shocks. The semiconductor shortages of 2021--2022 are reflected in our data as a contraction in newly established links, consistent with a period of strained production capacity and heightened uncertainty. 
The chip shortages in 2021--2022 also coincided with decreased centrality scores for firms in most regions.
One might naively have expected these crises to be associated with an increased dependency on a few firms (high closeness centrality) or the emergence of bottlenecks (high betweenness centrality).
However, the opposite is true, with both betweenness and closeness decreasing in 2021 and 2022 relative to the periods before and afterwards. 
The reasons for this are unclear and require further investigation.
One potential explanation is that the shortage disrupted existing hierarchical production structures, resulting in a more homogeneous distribution of centrality across firms. Potentially firms formed or reported relationships with other firms in the industry in a different way before and after the chip crises.

Second, The approach further allows us to trace how a shifting geopolitical landscape affects the network structure of the semiconductor industry. 
We observe a similar contraction in network link density in 2024-2025. We attribute this  contraction to intensified geopolitical tensions, particularly U.S.--China trade frictions.
A prominent driver of this trend has been the escalation of trade restrictions and export controls targeting the semiconductor sector. These policy shifts have incentivized, and in many cases forced, a strategic decoupling of supply chains between the two nations. This dynamic is clearly mirrored in our data by a pronounced restructuring of Chinese relationships in 2024 and 2025, where we observe a sharp collapse in links connecting Chinese firms to U.S. partners (see Fig.~\ref{fig:si-new-links}). These findings illustrate how web-derived networks can capture rapid structural adjustments in response to geopolitical shocks. Further, we can also track how geopolitical events affect individual firms. By examining how relationships with European, US, and Japanese firms evolve for selected lead companies, we uncover distinctive trajectories. This can shed light on how firms deal with an ever more dynamically evolving geopolitical landscape and improve the data basis for the rapidly evolving field of Geoeconomics \cite{mohr2025geoeconomics}.

Third, our results reveal structural shifts associated with major technological transitions. Most notably, our data set shows that NVIDIA strongly increases their network centrality measures starting in 2022, while TSMC's centrality is increasing from 2020 on. The timing coincides with the starting point of the generative AI boom, driven by the rapidly escalating computational demands required to train large language models and other foundation models. The release of GPT-3 in 2020 and the subsequent race toward ever-larger models, culminating in the widespread adoption of ChatGPT in 2022, generated an abrupt and highly specific surge in demand for parallel processing hardware, leading to a high demand in NVIDIA GPUs manufactured by TSMC. In parallel, NVIDIA became increasingly integrated both upstream and downstream in its supply chains, while also exhibiting a pronounced rise in betweenness centrality. Hence, our novel supply chain reconstruction method confirms NVIDIA's ascent to one of the nexus suppliers in the semiconductor industry \cite{shao2018data}.

\paragraph{Limitations and Granularity}
While our approach offers scalability and timeliness, it is subject to several limitations. First, reliance on Common Crawl introduces a coverage bias, as the reconstruction is restricted to firms with an informative and indexable web presence. Furthermore, the use of ORBIS to identify the initial population of semiconductor companies introduces an additional layer of selection bias; ORBIS tends to overrepresent larger firms and offers uneven coverage across different jurisdictions. It is possible that this database-level bias, favoring established, formal entities, is even more significant than the crawl-based bias itself. Consequently, our findings may understate the role of smaller, emerging players in the supply chain.

Second, an important aspect of interpreting the reconstructed network is its inherent incompleteness. Recent work based on firm-level VAT data provides the closest approximation to a census of real supply networks currently available, revealing extremely dense structures with average degrees between 30 and 50, alongside substantial churn at both the node (25--30\% year-on-year) and edge (around 55\% year-on-year) levels \cite{bacilieri2023firm,reisch2025supply}. Relative to these benchmarks, it is clear that our reconstruction only uncovers a small subset of the full network, as most links are not publicly declared. Our approach, therefore, recovers the observable portion of the network that is encoded in firms’ digital self-representations.

Third, the resulting network is unweighted and static in its link definitions. We lack granular data on link weights (e.g., transaction volumes or contract values) and the exact duration of these relationships, which limits our ability to distinguish between high-impact strategic alliances and minor supply agreements. Fourth, the reconstruction suffers from recall limitations. Our heuristic approach likely misses a large volume of existing relationships due to the absence of explicit web-based mentions or hyperlinks, meaning the network should be interpreted as a map of discoverable relationships rather than an exhaustive census.

Finally, resolving multinational entities remains challenging. Large conglomerates (e.g., Samsung) often rely on a single web domain (e.g., \texttt{samsung.com}) for their global operations, making it difficult to distinguish between national subsidiaries based solely on hyperlink structure. As a result, one-to-one matching between website edges and specific cross-border trade flows is constrained. To partially mitigate this, we approximate geographic attribution by matching domains to countries using revenue distribution data. Nevertheless, this remains a heuristic, and precise attribution of web-based links to specific cross-border transactions is inherently limited by the absence of distinct digital footprints for regional subsidiaries.

\section{Conclusion}

This study establishes that the open web, when processed at scale, serves as a vital repository of economic intelligence. By applying Large Language Models to millions of archived webpages, we have constructed a validated, dynamic multi-relational map of the global semiconductor supply chain that overcomes the latency and opacity of traditional administrative and business intelligence data, and importantly adds information that is complementary to these traditional data sources.
Our results confirm that this approach captures essential industry dynamics and tracks the impacts of changes in the geopolitical landscape and the ascent of AI. The reconstructed network accurately traces the contraction of supply links during the 2021 chip shortage and the rapid ascent of AI-specialized firms, such as NVIDIA, to central positions in the global ecosystem. These findings demonstrate that firm-level web data contains a robust economic signal that complements existing financial disclosures.
Ultimately, this framework offers a blueprint for a new approach to economic monitoring that can easily be applied to other sectors. As supply chains become increasingly complex and geopolitically sensitive, the ability to generate real-time, bottom-up visibility into inter-firm dependencies offers a crucial tool for policymakers and researchers aiming to assess resilience and navigate global shocks.

\section{Data and Methods}

\subsection{Data}

\subsubsection{Extracting relevant companies from ORBIS}

ORBIS~\citep{ORBIS_bvd} is a global database that compiles standardized information on over 460 million companies, including financials, industry classifications, ownership links, and website addresses. Following the methodology introduced in \cite{ASCII2024}, semiconductor companies were identified through a multi-step filtering process. First, companies listed in two industry-specific datasets (ETO and Abachy~\citep{georgetown-cset-2025-eto-chip-explorer, abachy-com-2024}) were matched to ORBIS to determine the relevant 4‑digit NACE industry codes. All NACE codes that occurred ten or more times among these matched companies were then used to extract all companies in ORBIS with these codes, yielding a pool of more than 800,000 potentially relevant firms. From this pool, only companies whose business descriptions contained at least one of the keywords “semiconductor,” “wafer,” “lithography,” “integrated circuit,” or “photomask” were retained. This resulted in 21,245 companies, which form the basis of our analysis. Companies without descriptive text in ORBIS were excluded. Among these, 8,538 had unique webpages and form the starting point for our web-based analysis.

\subsubsection{Common Crawl}

Common Crawl~\citep{commoncrawl} is a publicly available archive of web data that provides periodic snapshots of a large fraction of the World Wide Web. Using the company website information from ORBIS, we extracted archived pages from Common Crawl covering the period 2015–2025, with two snapshots per year, to build a temporal dataset. From these data, we take the induced subgraph of all semiconductor firms, i.e., considering links only between semiconductor firms and their nodes, enabling the construction of inter-company connection data over time. The technical details of data collection and orchestration are described in \cite{picatto2024dagster}.

While Common Crawl served as the primary data source, we identified that TSMC (Taiwan Semiconductor Manufacturing Company) was absent from the archive due to technical crawling anomalies, despite the absence of explicit exclusion protocols. Given TSMC's systemic importance as a central hub in the industry, we supplemented the dataset by independently scraping the company's press releases and news sections. 

\subsubsection{S\&P Dataset}

The S\&P Capital IQ dataset~\citep{spcapitaliq} (collected 2026), provides curated firm-level information on company fundamentals and company relationships such as: supply relationships, credits, partnerships, and ownership links, as reported in financial disclosures and company filings. We use it as an external benchmark for evaluating the plausibility and completeness of the reconstructed semiconductor production network derived from web data. Next to FactSet, S\&P Capital IQ is one of the typical business intelligence data sets used for international firm-level supply network analysis e.g., \citep{Chakraborty2024}.

The dataset covers 580,000 firms and 2Mio. unique directed relationships. The majority of relationships are of type supply (930,000) or credit (580,000), with the remainder comprising lease, distribution, licensing, and banking arrangements. Because we are interested in the flow of goods and services, we restrict our analysis to supply-type relationships.

To ensure a consistent comparison with the reconstructed semiconductor network, we restrict the S\&P Capital IQ dataset to firms operating in industries directly related to the semiconductor ecosystem. Specifically, we retain firms classified under the following primary industry categories:

\begin{itemize}
    \item Electronic Components
    \item Semiconductor Materials and Equipment
    \item Semiconductors
    \item Electronic Equipment and Instruments
    \item Technology Distributors
    \item Technology Hardware, Storage and Peripherals
    \item Communications Equipment
    \item Computer and Electronics Retail
    \item Electronic Manufacturing Services
\end{itemize}

This selection captures both upstream and downstream segments of the semiconductor value chain, including manufacturing, equipment supply, distribution, and end-product integration. This set of industries is broader than our keywords for filtering ORBIS, and allows us to retain relevant firms when semiconductors is not the primary industry of the company, e.g., for Samsung. The resulting subnetwork contains 18,000 firms and 48,671 directed supplier-costumer relationships.

\subsubsection{Validation Datasets}

To assess the plausibility of the reconstructed semiconductor network, we compare our results with two additional external datasets. First, we use the OECD Inter-Country Input–Output (ICIO) tables~\citep{oecdicio}, which capture sector-level trade and production linkages between countries. From the ICIO classification, we focus on the manufacturing sectors most relevant to semiconductors: ``C26'' (Computer, electronic and optical equipment), ``C27'' (Electrical equipment), and ``C28'' (Machinery and equipment, nec). Second, we employ international trade data from the BACI database~\citep{baci2020}, which provides highly disaggregated trade flows based on Harmonized System (HS) product codes. We select HS codes corresponding to wafer fabrication and semiconductor production. These two datasets serve as external benchmarks for validating the patterns and connectivity structures obtained from our reconstructed company-level network.

\subsection{Methods}

\subsubsection{Network construction}

\begin{figure}[H]
    \centering
    \includegraphics[width=\textwidth]{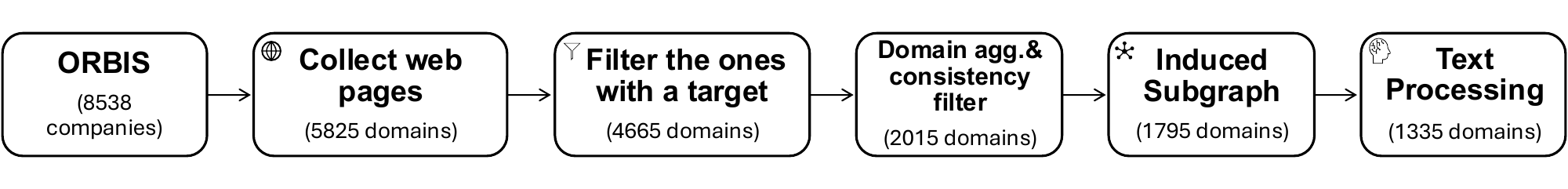}
    \caption{Methodological pipeline for reconstructing the firm-level network.}
    \label{fig:pipeline}
\end{figure}

The reconstruction of the semiconductor production network from web data followed a multi-stage pipeline involving webpage collection, filtering, domain aggregation, and automated relationship extraction (see Fig.~\ref{fig:pipeline} for an overview of the data reduction steps). Starting from the complete set of semiconductor firms identified in ORBIS, the main steps were as follows.

\paragraph{Preprocessing and cleaning.}
For each firm, we collected all archived webpages from the Common Crawl dataset (2015--2025). HTML files were parsed using \texttt{BeautifulSoup4} to extract plain text, and regular expressions were applied to remove boilerplate elements such as navigation menus, scripts, and duplicated layout components.  
We additionally excluded URLs containing terms such as \texttt{forum} or \texttt{community}, which correspond to user-generated content and were found to provide no relevant information on firm linkages.

\paragraph{Filtering irrelevant pages.}
Only a small fraction of company webpages contains information about other firms. We therefore applied two filtering steps to retain only potentially informative pages.  
First, we restricted the dataset to pages that link to at least one other domain from our semiconductor companies. This yielded the set of “source–target” domain pairs used in subsequent stages.  
Second, to avoid over-representing extremely common boilerplate sections (e.g., repetitive footer text), we limited the number of pages retained per source–target pair when mentions occurred excessively. This ensured that the remaining pages contained substantive textual content.

\paragraph{Domain aggregation and temporal consistency.}
Corporate web presence is heterogeneous: firms often maintain multiple regional or legacy domains (e.g., \texttt{company.com}, \texttt{company.co.kr}). We therefore aggregated such domains to a single parent entity to avoid artificial duplication of firms.  
Next, we applied a temporal consistency filter to select firms that appear reliably in the Common Crawl archive over the 2015--2025 period. This step ensured that the resulting panel reflects firms with stable archival coverage rather than those appearing sporadically due to crawl variability.  
Details of the consistency criteria and robustness checks, showing that including inconsistent domains only marginally increases network size, are provided in ~\ref{si:consistency}.

\paragraph{Induced subgraph construction.}
The temporally consistent set of firms defines the node set of the reconstructed network. Firms that passed the consistency filter but did not appear as either a source or target in any year naturally drop out of the induced graph. After applying all filtering and aggregation steps, the resulting network contains substantially fewer firms than the original ORBIS universe, reflecting the fact that only a subset of companies disclose inter-firm relationships on their publicly accessible webpages.

\paragraph{Keyword-based selection.}
To further refine the dataset, we searched each retained webpage for supply-chain-related terminology. This consisted of the terms \texttt{supply}, \texttt{vendor}, \texttt{procure}, \texttt{manufacture}, \texttt{produce}, \texttt{assemble}, \texttt{purchase}, \texttt{client}, \texttt{acquire}, \texttt{partner}, \texttt{collaborate}, \texttt{cooperate}, \texttt{alliance}, \texttt{joint venture}, and \texttt{strategic alliance}. Only pages containing at least one such keyword were sent to the relationship extraction stage. This step proved permissive, as the vast majority of retained pages already contained at least one of the specified terms, resulting in only a marginal reduction in the total number of pages sent to the LLM stage. 

\paragraph{Relationship extraction using large language models.}
The filtered pages were processed using the LLM GPT-4o-mini to infer the type and direction of inter-firm relationships. Company webpages vary widely in format, ranging from sparse tables and product catalogues to multilingual descriptions, so the model was instructed to extract relations even in the absence of structured narratives.  
For each webpage, the model received the source domain, the referenced target domain, and the cleaned page text, along with the following instruction:

\begin{quote}
You are a supply chain relationship extraction expert analyzing the webpage of \{src\_domain\}. In this page, \{dst\_domain\} is mentioned. Your goal is to infer the type and direction of any business relationship between \{src\_domain\} and \{dst\_domain\} based on the provided text.

Consider both directions equally. Do not assume a default unless explicitly stated.

- If it is clear from the text that \{src\_domain\} supplies/manufactures to \{dst\_domain\} (Return 1) 

- If it is clear from the text that \{dst\_domain\} supplies/manufactures to \{src\_domain\} (Return 2)

- If it is clear from the text that \{src\_domain\} is only partners/collaborators with \{dst\_domain\} without a clear supplier/buyer direction (Return 3)

- If it is clear from the text that \{src\_domain\} owns/acquired \{dst\_domain\} (Return 4)

- If it is clear from the text that \{dst\_domain\} owns/acquired \{src\_domain\} (Return 5)

If no relationship is found or the text is not clear enough, return 0.
\end{quote}

The model returned a single digit for each source–target pair. A total of approximately 200,000 pages were processed in batches, costing around 130 USD for all crawls.

This pipeline produced a directed network of inter-company relationships, distinguishing between supplier–buyer links, partnerships, and acquisitions.

\subsubsection{Network validation}

\paragraph{Validation of LLM Link-Extraction Performance.}

To evaluate the performance of the GPT-based relationship extraction, 
we conducted a manual validation study. 
A random sample of 100 webpages was selected from the processed dataset, 
ensuring a balanced representation so that pages with no identified relationships 
did not dominate the sample. 
Each page was manually annotated by multiple annotators to determine 
(i) whether a supply chain relationship was present (binary classification) and 
(ii) if present, the type and direction of the relationship (multi-class classification). These annotations were then compared against the GPT model outputs. 

\paragraph{Country Level Validation.}

To assess the plausibility of the firm-level network reconstructed from websites, 
we performed a soft validation by aggregating firm-level relationships to the 
country level and comparing the resulting structure with external datasets 
(OECD ICIO and BACI trade data).

Each domain in our firm-level network was mapped to one or more countries using ORBIS data on company headquarters and revenue. When a company reported activity in multiple countries, we used the reported revenue shares to proportionally distribute its outgoing and incoming links across those countries. For every firm-level edge $(u, v)$ linking firms $u$ and $v$, we combined the country-level revenue shares and magnitudes to assign weights to the corresponding country--country pairs. Specifically, for each pair of countries $(c_1, c_2)$ associated with the two firms, the contribution to the country-level edge weight was defined as
\[
w_{c_1 c_2}^{(u,v)} = \log\!\big(1 + R_u R_v\big) \, s_{u,c_1} \, s_{v,c_2},
\]
where $R_u$ and $R_v$ denote the total revenues of firms $u$ and $v$, and $s_{u,c_1}$ and $s_{v,c_2}$ are their respective revenue shares in countries $c_1$ and $c_2$. The overall country-level weight was then obtained by summing these contributions across all firm-level links:
\[
W_{c_1 c_2} = \sum_{(u,v) \in E} w_{c_1 c_2}^{(u,v)}.
\]
This yields a weighted and directed country-level network, where edge weights reflect the intensity of cross-country relationships inferred from firm-level connections and revenue distributions.

\paragraph{Firm-Level Validation}
\label{conf_model}

We validate the reconstructed semiconductor production network by comparison with an independent commercial dataset from S\&P Capital IQ, which provides curated information on directed supplier--customer relationships between firms. While S\&P Capital IQ is high-quality, it is not exhaustive: its coverage is biased toward large and publicly visible firms and underrepresents smaller or private actors. The reconstructed web-based network is likewise incomplete and subject to different sampling biases. Therefore, neither dataset can be treated as ground truth; instead, the comparison evaluates whether two independent observations of the same underlying system exhibit statistically significant structural agreement.

To align firms across datasets, we map company names from S\&P Capital IQ to web domains used in the reconstructed network. Because large corporations often operate through multiple legal entities (e.g., subsidiaries across regions), a direct name-to-name matching would artificially fragment the network.

To address this, we aggregate firms at the \textbf{domain level}: all companies associated with the same web domain are merged into a single node. This effectively consolidates subsidiaries and regional entities into their parent organization when they share a common digital presence.

For example, multiple legal entities such as Sony Semiconductor Solutions Corporation, Sony Mobile Communications (USA) Inc., and Sony Semiconductor Manufacturing Corporation are aggregated under a single node corresponding to the \texttt{sony} domain. 

This aggregation ensures consistency between datasets and prevents overcounting of nodes and edges due to corporate fragmentation.

To ensure comparability, we restrict attention to \textbf{directed supplier $\rightarrow$ customer links}. From the reconstructed network, we retain only edges classified as supply-chain relationships. 

The overlap between the two datasets is evaluated at both the node and edge levels. While a substantial fraction of firms in the reconstructed network are also present in S\&P, simple overlap counts are not sufficient to assess significance, as highly connected firms are more likely to appear in both datasets.

To control for this effect, we compare the observed overlap against a \textbf{directed configuration model} defined on the set of shared firms. This model preserves the in- and out-degree sequences of each dataset while randomizing connections, thereby providing a null hypothesis in which overlap arises solely due to degree heterogeneity.

Let $k_i^{out,A}$ and $k_i^{in,A}$ denote the out- and in-degree of firm $i$ in dataset $A$ (S\&P), and analogously for dataset $B$ (reconstructed network). Let $m_A$ and $m_B$ denote the total number of directed edges in each dataset restricted to shared firms. Under the configuration model, the probability of a directed edge from $i$ to $j$ is approximated by

\begin{equation}
P_{ij}^A \approx \frac{k_i^{out,A} \, k_j^{in,A}}{m_A},
\end{equation}

and similarly for dataset $B$. The expected number of overlapping directed edges is then

\begin{equation}
\mathbb{E}[m_{\cap}] =
\frac{1}{m_A m_B}
\sum_{i \neq j}
\left(k_i^{out,A} k_i^{out,B}\right)
\left(k_j^{in,A} k_j^{in,B}\right).
\end{equation}

Statistical significance is assessed via $z$-scores computed from the distribution of overlaps generated by repeated randomizations under the configuration model.

To further understand the structural origin of the overlap, we partition firms into core and periphery groups using the core--periphery structure identified in the reconstructed network. We then decompose edges into three categories: core--core, core--periphery, and periphery--periphery.

For each category, we compute the observed overlap and compare it to the corresponding expectation under the configuration model, restricted to edges within that subset. This allows us to assess whether the agreement between datasets is driven primarily by highly connected core firms or reflects broader structural consistency across the network.

To assess statistical significance within the core--periphery decomposition, we estimate the null distribution using Monte Carlo simulations of the directed configuration model. Specifically, we generate 1,000 randomized networks via degree-preserving edge rewiring on the set of shared nodes and recompute overlap counts within each edge category (core--core, core--periphery, periphery--periphery).

\section*{Data Availability}
The datasets analyzed during the current study comprise a combination of public and proprietary sources. The public datasets are openly available from their respective providers: Common Crawl (\url{https://commoncrawl.org}), the OECD Inter-Country Input-Output (ICIO) Tables (\url{https://www.oecd.org/en/data/datasets/inter-country-input-output-tables.html}), and the BACI international trade database provided by CEPII (\url{https://www.cepii.fr/CEPII/en/bdd_modele/bdd_modele_item.asp?id=37}).

The proprietary financial datasets (ORBIS and S\&P Capital IQ) were used under commercial license for the current study and are not publicly available due to third-party restrictions. 

All aggregated, derived data that support the findings of this study are available from the corresponding author upon reasonable request.

\section*{Code Availability} Code is available upon request directly from the authors.

\bibliographystyle{unsrt} 
\bibliography{references}

\begin{thebibliography}{10}

\bibitem{pichler2023sci}
Anton Pichler, Christian Diem, Alexandra Brintrup, François Lafond, Glenn Magerman, Gert Buiten, Thomas~Y. Choi, Vasco~M. Carvalho, J.~Doyne Farmer, and Stefan Thurner.
\newblock Building an alliance to map global supply networks.
\newblock {\em Science}, 382(6668):270--272, 2023.

\bibitem{acemoglu2012network}
Daron Acemoglu, Vasco~M Carvalho, Asuman Ozdaglar, and Alireza Tahbaz-Salehi.
\newblock The network origins of aggregate fluctuations.
\newblock {\em Econometrica}, 80(5):1977--2016, 2012.

\bibitem{cimini2015systemic}
Giulio Cimini, Tiziano Squartini, Diego Garlaschelli, and Andrea Gabrielli.
\newblock Systemic risk analysis on reconstructed economic and financial networks.
\newblock {\em Scientific reports}, 5(1):15758, 2015.

\bibitem{klimek2019resil}
Peter Klimek and Sebastian Poledna.
\newblock Quantifying economic resilience from input–output susceptibility to improve predictions of economic growth and recovery.
\newblock {\em Nature Communications}, 10(1):1677, 2019.

\bibitem{inoue2019firm}
Hiroyasu Inoue and Yasuyuki Todo.
\newblock Firm-level propagation of shocks through supply-chain networks.
\newblock {\em Nature Sustainability}, 2(9):841--847, 2019.

\bibitem{acemoglu2011}
Daron Acemoglu, Vasco Carvalho, Asuman Ozdaglar, and Alireza Tahbaz-Salehi.
\newblock The network origins of aggregate fluctuations.
\newblock {\em Econometrica}, 80, 10 2011.

\bibitem{carvalho21}
Vasco~M Carvalho, Makoto Nirei, Yukiko~U Saito, and Alireza Tahbaz-Salehi.
\newblock Supply chain disruptions: Evidence from the great east {J}apan earthquake*.
\newblock {\em The Quarterly Journal of Economics}, 136(2):1255--1321, 12 2020.

\bibitem{diem2024}
Christian Diem, András Borsos, Tobias Reisch, János Kertész, and Stefan Thurner.
\newblock Estimating the loss of economic predictability from aggregating firm-level production networks.
\newblock {\em PNAS Nexus}, 3(3):pgae064, 02 2024.

\bibitem{Chakraborty2024}
Abhijit Chakraborty, Tobias Reisch, Christian Diem, Pablo Astudillo-Estévez, and Stefan Thurner.
\newblock Inequality in economic shock exposures across the global firm-level supply network.
\newblock {\em Nature Communications}, 15(1):3348, 2024.

\bibitem{papadopoulos2025climate}
Georgios Papadopoulos, Javier Ojea~Ferreiro, and Roberto Panzica.
\newblock Climate stress test of the global supply chain network: the case of river floods.
\newblock 2025.

\bibitem{diem2022quantifying}
Christian Diem, Andr{\'a}s Borsos, Tobias Reisch, J{\'a}nos Kert{\'e}sz, and Stefan Thurner.
\newblock Quantifying firm-level economic systemic risk from nation-wide supply networks.
\newblock {\em Scientific reports}, 12(1):7719, 2022.

\bibitem{mancini2025evolution}
Anna Mancini, Bal\'azs Lengyel, Riccardo Di~Clemente, and Giulio Cimini.
\newblock Evolution and determinants of firm-level systemic risk in local production networks.
\newblock {\em arXiv preprint arXiv:2506.21426}, 2025.

\bibitem{Diem03052025}
Christian Diem, William Schueller, Melanie Gerschberger, Johannes Stangl, Beate Conrady, Markus Gerschberger, and Stefan Thurner.
\newblock Supply network stress-testing of food security on the establishment-level.
\newblock {\em International Journal of Production Research}, 63(9):3259--3283, 2025.

\bibitem{McNerney2022}
James McNerney, Charles Savoie, Francesco Caravelli, Vasco~M. Carvalho, and J.~Doyne Farmer.
\newblock How production networks amplify economic growth.
\newblock {\em Proceedings of the National Academy of Sciences}, 119(1):e2106031118, 2022.

\bibitem{vitali2011network}
Stefania Vitali, James~B Glattfelder, and Stefano Battiston.
\newblock The network of global corporate control.
\newblock {\em PloS one}, 6(10):e25995, 2011.

\bibitem{mizuno2020power}
Takayuki Mizuno, Shohei Doi, and Shuhei Kurizaki.
\newblock The power of corporate control in the global ownership network.
\newblock {\em Plos one}, 15(8):e0237862, 2020.

\bibitem{dahlke2024epidemic}
Johannes Dahlke, Mathias Beck, Jan Kinne, David Lenz, Robert Dehghan, Martin W{\"o}rter, and Bernd Ebersberger.
\newblock Epidemic effects in the diffusion of emerging digital technologies: evidence from artificial intelligence adoption.
\newblock {\em Research Policy}, 53(2):104917, 2024.

\bibitem{zhu2006innovation}
Kevin Zhu, Shutao Dong, Sean~Xin Xu, and Kenneth~L Kraemer.
\newblock Innovation diffusion in global contexts: determinants of post-adoption digital transformation of {E}uropean companies.
\newblock {\em European journal of information systems}, 15(6):601--616, 2006.

\bibitem{economist_nvidia_openai_2025}
{The Economist}.
\newblock Nvidia's \$100bn bet on {OpenAI} raises more questions than it answers.
\newblock {\em The Economist}, September 2025.
\newblock Business section. Published September 22, 2025. Accessed 2026-01-09.

\bibitem{ASCII2024}
Peter Klimek, Markus Gerschberger, Christopher Schwarz, Tiberiu-Alexandru Cioban, Agnes Kügler, Elma Dervic, Georg Heiler, Hernan Picatto, Klaus Friesenbichler, and Lukas Schmoigl.
\newblock Mapping of the global semiconductor supply chain: Embedding {A}ustria in the global semiconductor inter-firm network.
\newblock Policy brief, Supply Chain Intelligence Institute Austria (ASCII), May 2024.

\bibitem{oecd2025mappingsemiconductor}
{OECD}.
\newblock Mapping the semiconductor value chain: Working towards identifying dependencies and vulnerabilities.
\newblock Technical Report 182, OECD Publishing, Paris, June 2025.

\bibitem{bacilieri2023firm}
Andrea Bacilieri, Andr{\'a}s Borsos, Pablo Astudillo-Estevez, and Fran{\c{c}}ois Lafond.
\newblock Firm-level production networks: what do we (really) know.
\newblock {\em INET Oxford Working Paper}, 2023, 2023.

\bibitem{Dietzenbacher01032013}
Erik Dietzenbacher, Bart Los, Robert Stehrer, Marcel Timmer, and Gaaitzen de~Vries.
\newblock The construction of world input–output tables in the wiod project.
\newblock {\em Economic Systems Research}, 25(1):71--98, 2013.

\bibitem{dhyne2015belgian}
Emmanuel Dhyne, Glenn Magerman, and Stela Rub{\'\i}nov{\'a}.
\newblock The {B}elgian production network 2002-2012.
\newblock Technical report, NBB Working Paper, 2015.

\bibitem{mungo2023reconstructing}
Luca Mungo, Fran{\c{c}}ois Lafond, Pablo Astudillo-Est{\'e}vez, and J~Doyne Farmer.
\newblock Reconstructing production networks using machine learning.
\newblock {\em Journal of Economic Dynamics and Control}, 148:104607, 2023.

\bibitem{maccarthy2022mapping}
Bart~L MacCarthy, Wafaa~AH Ahmed, and Guven Demirel.
\newblock Mapping the supply chain: Why, what and how?
\newblock {\em International Journal of Production Economics}, 250:108688, 2022.

\bibitem{wu2021analysis}
Xiling Wu, Caihua Zhang, and Wei Du.
\newblock An analysis on the crisis of “chips shortage” in automobile industry——based on the double influence of {COVID}-19 and trade friction.
\newblock In {\em Journal of Physics: Conference Series}, volume 1971, page 012100. IOP Publishing, 2021.

\bibitem{mohammad2022global}
Wassen Mohammad, Adel Elomri, and Laoucine Kerbache.
\newblock The global semiconductor chip shortage: Causes, implications, and potential remedies.
\newblock {\em IFAC-PapersOnLine}, 55(10):476--483, 2022.

\bibitem{bednarski2025geopolitical}
Lukasz Bednarski, Samuel Roscoe, Constantin Blome, and Martin~C Schleper.
\newblock Geopolitical disruptions in global supply chains: a state-of-the-art literature review.
\newblock {\em Production planning \& control}, 36(4):536--562, 2025.

\bibitem{mungo2024reconstructing}
Luca Mungo, Alexandra Brintrup, Diego Garlaschelli, and Fran{\c{c}}ois Lafond.
\newblock Reconstructing supply networks.
\newblock {\em Journal of Physics: Complexity}, 5(1):012001, 2024.

\bibitem{ialongo2022reconstructing}
Leonardo~Niccol{\`o} Ialongo, Camille De~Valk, Emiliano Marchese, Fabian Jansen, Hicham Zmarrou, Tiziano Squartini, and Diego Garlaschelli.
\newblock Reconstructing firm-level interactions in the dutch input--output network from production constraints.
\newblock {\em Scientific reports}, 12(1):11847, 2022.

\bibitem{wichmann2020extracting}
P~Wichmann, A~Brintrup, S~Baker, P~Woodall, and D~McFarlane.
\newblock Extracting supply chain maps from news articles using deep neural networks.
\newblock {\em International Journal of Production Research}, 58(17):5320--5336, 2020.

\bibitem{almahri2024enhancing}
Sara AlMahri, Liming Xu, and Alexandra Brintrup.
\newblock Enhancing supply chain visibility with knowledge graphs and large language models.
\newblock {\em arXiv preprint arXiv:2408.07705}, 2024.

\bibitem{jackson2025supply}
Ilya Jackson, Maria Jes{\'u}s~Sa{\'e}nz, Dmitry Ivanov, and Benedict~Jun Ma.
\newblock Supply chain mapping through retrieval-augmented generation: applications to the electronics industry.
\newblock {\em Journal of the Operational Research Society}, pages 1--21, 2025.

\bibitem{fujiwara2010large}
Yoshi Fujiwara and Hideaki Aoyama.
\newblock Large-scale structure of a nation-wide production network.
\newblock {\em The European Physical Journal B}, 77(4):565--580, 2010.

\bibitem{brintrup2016topological}
Alexandra Brintrup, Anna Ledwoch, and Jose Barros.
\newblock Topological robustness of the global automotive industry.
\newblock {\em Logistics Research}, 9(1):1, 2016.

\bibitem{mohr2025geoeconomics}
Cathrin Mohr and Christoph Trebesch.
\newblock Geoeconomics.
\newblock {\em Annual Review of Economics}, 17, 2025.

\bibitem{shao2018data}
Benjamin~BM Shao, Zhan~Michael Shi, Thomas~Y Choi, and Sangho Chae.
\newblock A data-analytics approach to identifying hidden critical suppliers in supply networks: Development of nexus supplier index.
\newblock {\em Decision Support Systems}, 114:37--48, 2018.

\bibitem{reisch2025supply}
Tobias Reisch, Andr{\'a}s Borsos, and Stefan Thurner.
\newblock Supply chain network rewiring dynamics at the firm-level.
\newblock {\em arXiv preprint arXiv:2503.20594}, 2025.

\bibitem{ORBIS_bvd}
{Bureau van Dijk}.
\newblock Orbis database.
\newblock \url{https://www.bvdinfo.com/en-gb/our-products/data/orbis}.
\newblock Accessed 2025-11-10.

\bibitem{georgetown-cset-2025-eto-chip-explorer}
{Georgetown CSET}.
\newblock Eto chip explorer.
\newblock \url{https://github.com/georgetown-cset/eto-chip-explorer}.
\newblock Accessed 2025-11-10.

\bibitem{abachy-com-2024}
{Abachy.com}.
\newblock Semiconductor materials and equipment.
\newblock \url{https://abachy.com/}.
\newblock Accessed 2025-11-10.

\bibitem{commoncrawl}
{Common Crawl Foundation}.
\newblock Common crawl dataset.
\newblock \url{https://commoncrawl.org/}.
\newblock Accessed 2025-11-10.

\bibitem{picatto2024dagster}
Hernan Picatto, Georg Heiler, and Peter Klimek.
\newblock Cost-effective big data orchestration using dagster: A multi-platform approach.
\newblock {\em arXiv preprint arXiv: 2408.11635}, 2024.

\bibitem{spcapitaliq}
{S\&P Global Market Intelligence}.
\newblock S\&{P} capital {IQ} database.
\newblock \url{https://www.capitaliq.com}.
\newblock Accessed 2025-11-10.

\bibitem{oecdicio}
{OECD}.
\newblock Inter-country input–output ({ICIO}) tables.
\newblock \url{https://www.oecd.org/sti/ind/inter-country-input-output-tables.htm}, 2023.
\newblock Accessed 2025-11-10.

\bibitem{baci2020}
Guillaume Gaulier and Soledad Zignago.
\newblock Baci: International trade database at the product-level (the 2020 version).
\newblock \url{https://www.cepii.fr/CEPII/en/bdd_modele/presentation.asp?id=37}, 2020.
\newblock Accessed 2025-11-10.

\end{thebibliography}

\newpage
\appendix
\setcounter{section}{0}
\renewcommand{\thesection}{S\arabic{section}}
\renewcommand{\theHsection}{S\arabic{section}} 

\setcounter{figure}{0}
\renewcommand{\thefigure}{S\arabic{figure}}
\renewcommand{\theHfigure}{S\arabic{figure}} 

\setcounter{table}{0}
\renewcommand{\thetable}{S\arabic{table}}
\renewcommand{\theHtable}{S\arabic{table}}

\begin{center}
    \Large\textbf{Supplementary Information}
\end{center}

\setcounter{section}{0}
\renewcommand{\thesection}{S\arabic{section}}

\counterwithin{figure}{section}
\renewcommand{\thefigure}{\thesection.\arabic{figure}}

\section{Consistency Filtering and Robustness Analysis}
\label{si:consistency}

To ensure the temporal reliability of domains used in our analysis, we constructed a binary activity string for each domain indicating the years in which it appeared. We then applied a set of consistency rules designed to remove domains whose temporal patterns suggested noise, crawler artefacts, or unstable hosting. A domain was retained if it satisfied at least one of the following criteria:

\begin{enumerate}
    \item The activity string ended with at least three consecutive 1's (\texttt{111}), ensuring retention of domains consistently active in the later years.
    \item The total number of 1's (active years) exceeded four.
    \item The activity string contained at least one sequence of three or more consecutive 1's.
    \item The activity string did not contain any sequence of three or more consecutive 0's, except in cases where the pattern ended in \texttt{111}.
\end{enumerate}

Applying these criteria yielded a core set of 2015 temporally consistent domains.  
These domains were used in the main analysis.

\subsection{Robustness Check Using a Broader Domain Set}

To assess the robustness of our results, we repeated the graph construction using a more inclusive set of all domains that appeared in at least two distinct years (3699 domains in total). This represents a \textbf{+56\%} increase in the number of domains relative to the consistent set.

The unprocessed supplier--customer graph derived from this broader set contained 2923 nodes and 18{,}110 raw edges, whereas the graph derived from the 2015 consistent domains contained 1795 nodes and 16{,}225 raw edges.

Although relaxing the filtering expanded the domain pool substantially (\textbf{+84\%} more domains), the resulting graph exhibited only an \textbf{11.6\%} increase in raw edges (from 16{,}225 to 18{,}110). In other words, adding 1128 additional domains contributed only 1885 additional edges, indicating that most of these domains were either peripheral or sparsely connected.

\medskip

\noindent These percentages demonstrate that the majority of meaningful relational structure in the network is already captured by the temporally consistent subset. The broader domain set changes the size of the graph only modestly, supporting the robustness of our results with respect to the filtering criteria.

\section{Validation}

\begin{figure}[ht]
    \centering
    \includegraphics[width=1\textwidth]{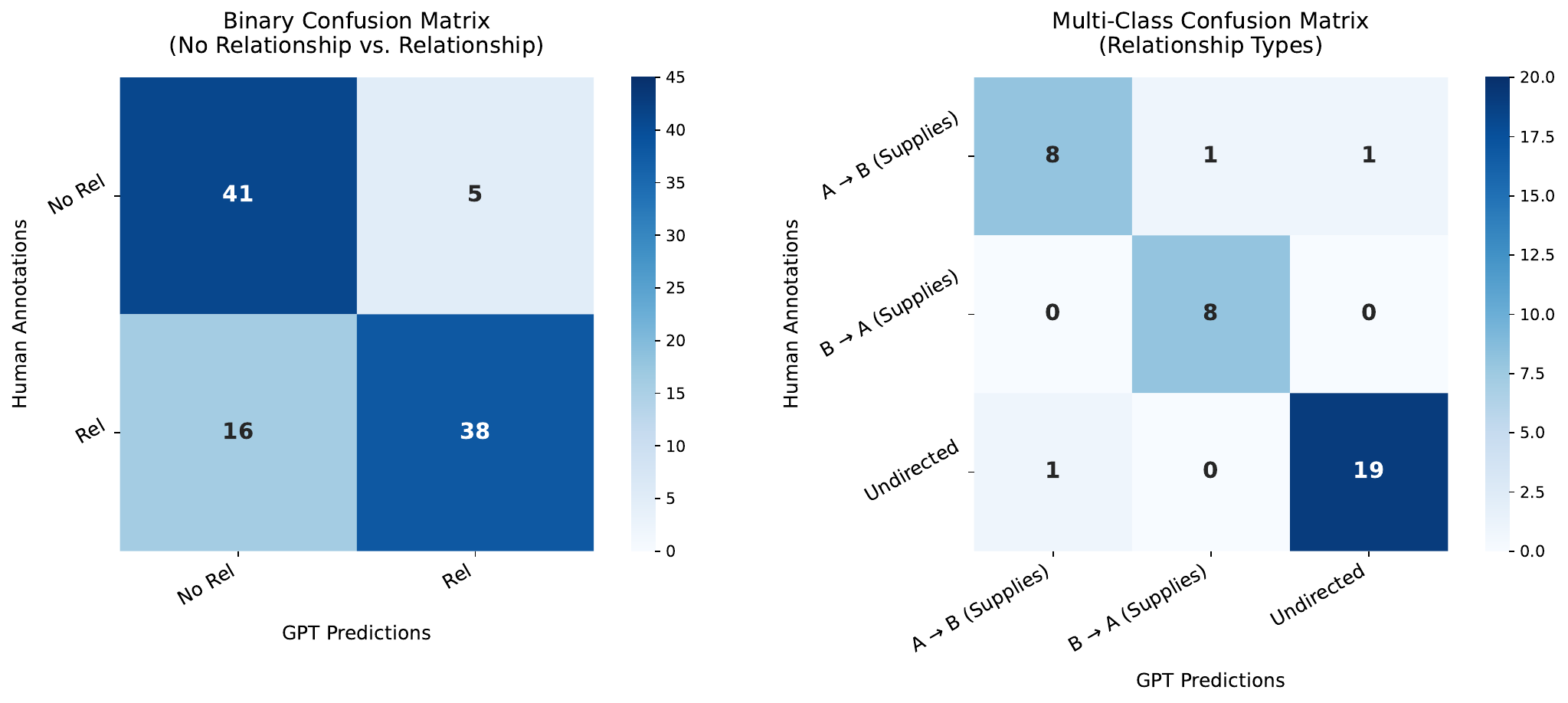}
    \caption{Confusion matrix heatmap for the GPT-based relationship classification.}
    \label{fig:relationship_validation}
\end{figure}

\begin{figure}[htbp]
    \centering
    \includegraphics[width=\textwidth]{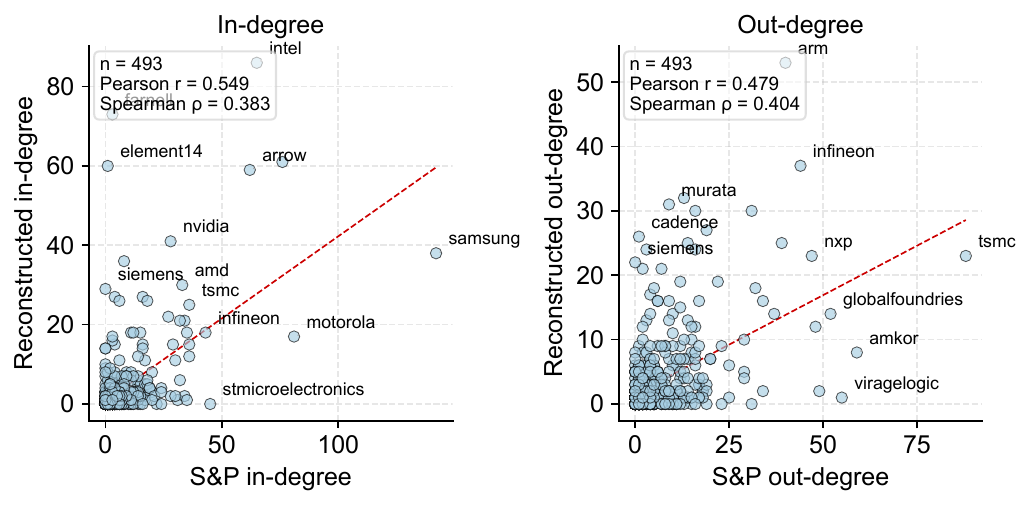}
    \caption{\textbf{Comparison of firm-level connectivity between the S\&P network and the reconstructed Common Crawl graph.}
    The scatter plots compare the \textbf{(a)} in-degree and \textbf{(b)} out-degree for the 493 overlapping domains identified in both datasets.
    We observe positive degree correlations between the two sources: for in-degree (suppliers), Pearson $r=0.549$ and Spearman $\rho=0.383$; for out-degree (customers), Pearson $r=0.479$ and Spearman $\rho=0.404$.
    To ensure a consistent comparison, regional subsidiaries in the S\&P dataset were aggregated into single global entities to match the domain-based granularity of the Common Crawl data.
    The red dashed line indicates the best linear fit through the scatter points (regression line).}
    \label{fig:degree_comparison}
\end{figure}

\subsection{Graph Overlap Significance}

To test whether the observed edge overlap between the S\&P and Common Crawl 
(CC) supply chain graphs exceeds what is expected given the degree sequences 
alone, we assess significance via Monte Carlo rewiring with 1,000 
degree-preserving rewirings of the CC graph. All $p$-values are bounded 
below by $1/N_\text{iter} = 10^{-3}$.

\begin{table}[h]
\centering
\caption{Edge overlap significance — full induced subgraph on common nodes.}
\label{tab:overlap_full}
\begin{tabular}{lccc}
\toprule
Mode & Observed & $\mathbb{E}[m_\cap]$ & $z$ \\
\midrule
Undirected (all CC edge types) & 682 & 415.40 & 22.02 \\
Directed (supply edges only)   & 380 & 165.30 & 23.14 \\
\bottomrule
\end{tabular}
\end{table}

\subsection*{K-core Tier Analysis}

We decompose the common nodes using $k$-core decomposition with threshold 
$k^* = 10$, yielding 156 core and 431 periphery nodes in the undirected 
case (125 and 368 in the directed case). Edges are partitioned into 
core--core, core--periphery, and periphery--periphery groups and the 
configuration model test is run separately on each.

\begin{table}[h]
\centering
\caption{Edge overlap significance by k-core tier ($k^* = 10$, 
         Monte Carlo with 1,000 rewirings).}
\label{tab:overlap_tiers}
\begin{tabular}{llccccc}
\toprule
Mode & Group & S\&P edges & CC edges & Observed & $\mathbb{E}[m_\cap]$ & $z$ \\
\midrule
\multirow{3}{*}{Undirected}
  & Core--core           & 1,689 &   949 & 404 & 270.28 & 13.92 \\
  & Core--periphery      & 1,155 & 1,257 & 206 &  26.45 & 37.45 \\
  & Periphery--periphery &   409 &   493 &  72 &   3.06 & 42.38 \\
\midrule
\multirow{3}{*}{Directed}
  & Core--core           & 1,498 &   552 & 221 & 137.72 & 10.98 \\
  & Core--periphery      & 1,086 &   790 & 119 &  30.86 & 22.67 \\
  & Periphery--periphery &   447 &   350 &  40 &  14.13 & 15.01 \\
\bottomrule
\end{tabular}
\end{table}

\noindent All six groups are statistically significant ($p < 10^{-3}$). 
$z$-scores should be used for cross-group comparison as all $p$-values 
hit the Monte Carlo floor of $1/N_\text{iter}$.

\section{Country Level Analysis}

\begin{figure}[H]
    \centering
    \subfloat[Comparison with OECD ICIO matrix.]{
        \includegraphics[width=\textwidth]{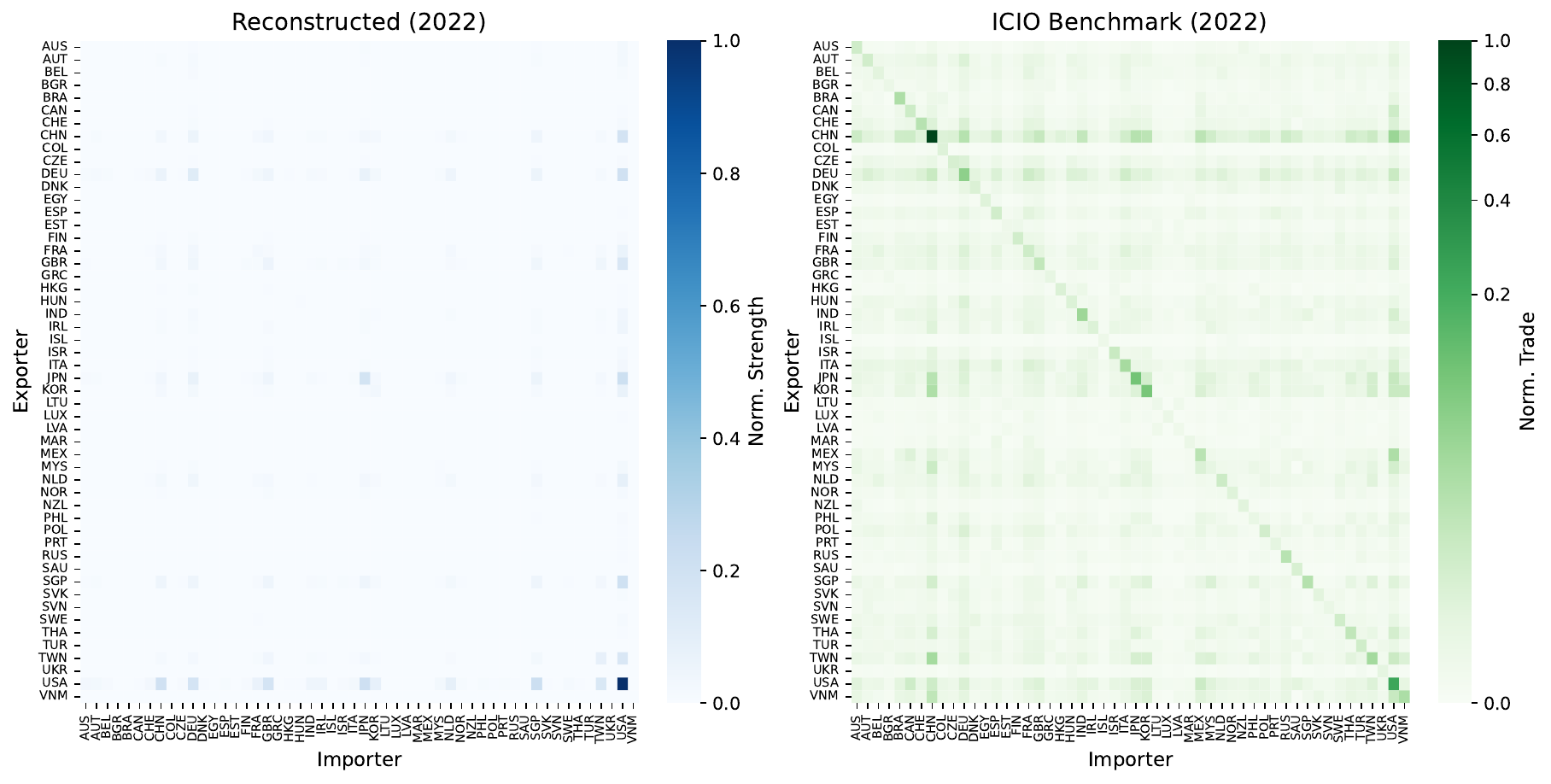}
    }

    \vspace{0.4cm}

    \subfloat[Comparison with BACI trade flows.]{
        \includegraphics[width=\textwidth]{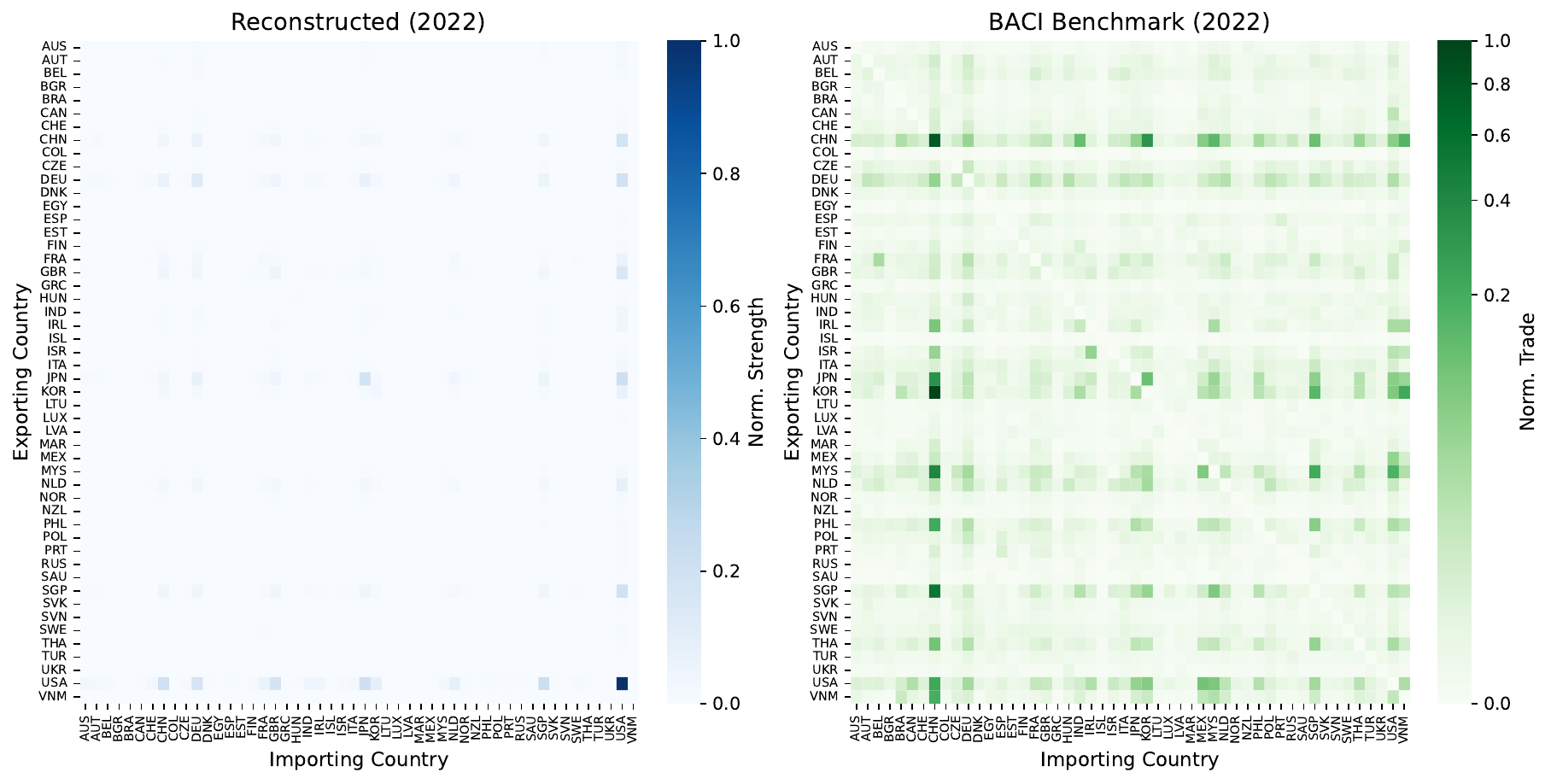}
    }

    \caption{Heatmaps of the aggregated country-level network compared with external datasets.}
    \label{fig:soft_validation}
\end{figure}

\begin{figure}[ht]
    \centering
    \includegraphics[width=\textwidth]{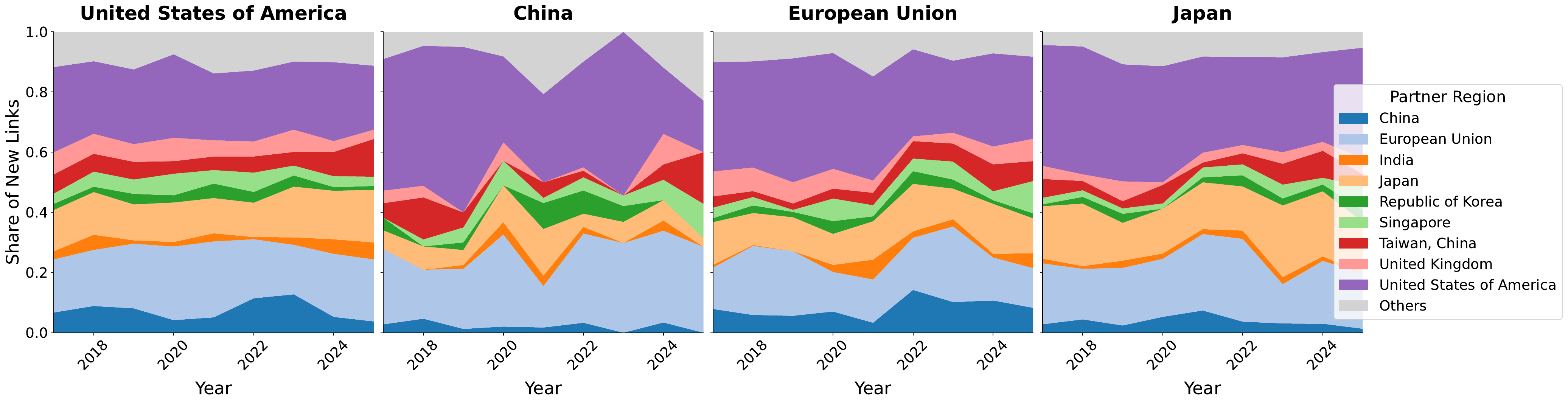}
    \caption{Share of newly observed inter-firm links by partner country, 2017–2024. Each panel shows one focal economy’s distribution of new connections with major trading partners.}
    \label{fig:si-new-links}
\end{figure}

\end{document}